\documentclass[10pt,english]{article}
\usepackage{graphicx}

\begin{document}

\title{Phase transition in a class of \\ non-linear random networks}

\author{M. Andrecut$^{1}$ and S. A. Kauffman$^{2}$}

\maketitle
{

\centering $^{1}$Institute for Space Imaging Science \\ $^{2}$Institute for Biocomplexity and Informatics 

}{

\centering University of Calgary, Alberta, T2N 1N4, Canada 

}

\begin{abstract}
We discuss the complex dynamics of a non-linear random networks model,
as a function of the connectivity $k$ between the elements of the
network. We show that this class of networks exhibit an order-chaos
phase transition for a critical connectivity $k_{c}=2$. Also, we
show that both, pairwise correlation and complexity measures are maximized
in dynamically critical networks. These results are in good agreement
with the previously reported studies on random Boolean networks and
random threshold networks, and show once again that critical networks
provide an optimal coordination of diverse behavior. 
\end{abstract}

\section{Introduction}

Random Boolean networks (RBNs) are a class of complex systems, that
show a well-studied transition between ordered and disordered phases.
The RBN model was initially introduced as an idealization of genetic
regulatory networks. Since then, the RBN model has attracted much
interest in a wide variety of fields, ranging from cell differentiation
and evolution to social and physical spin systems (for a review of
the RBN model see \cite{key-1} and \cite{key-2}, and the references within).
The dynamics of RBNs can be classified as ordered, disordered, or
critical, as a function of the average connectivity $k$, between
the elements of the network, and the bias $p$ in the choice of Boolean
functions. For equiprobable Boolean functions, $p=1/2$, the critical
connectivity is $k_{c}=2$. The RBNs operating in the ordered regime
$(k<k_{c})$ exhibit simple dynamics, and are intrinsically robust
under structural and transient perturbations. In contrast, the RBNs
in the disordered regime $(k>k_{c})$ are extremely sensitive to small
perturbations, which rapidly propagate throughout the entire system.
Recently, it has been shown that the pairwise mutual information exhibits
a jump discontinuity at the critical value $k_{c}$ of the RBN model
\cite{key-3}. More recently, similar results have been reported for a related
class of discrete dynamical networks, called random threshold networks
(RTNs) \cite{key-4}.

In this paper we consider a non-linear random networks (NLRNs) model,
which represents a departure from the discrete valued state representation,
corresponding to the RBN and RTN models, to a continuous valued state
representation. We discuss the complex dynamics of the NLRN model,
as a function of the average connectivity (in-degree) $k$. We show
that the NLRN model exhibits an order-chaos phase transition, for
the same critical connectivity value $k_{c}=2$, as the RBN and RTN
models. Also, we show that both, pairwise correlation and complexity
measures are maximized in dynamically critical networks. These results
are in very good agreement with the previously reported studies on
the RBN and RTN models, and show once again that critical networks
provide an optimal coordination of diverse behavior.

\section{NLRN model}

The NLRN model consists of $N$ randomly interconnected variables,
with continuously valued states $-1\leq x_{n}\leq+1$, $n=1,...,N$.
At time $t$ the state of the network is described by an $N$ dimensional
vector\begin{equation}
\mathbf{x}(t)=[x_{1}(t),...,x_{N}(t)]^{T},\end{equation}
 which is updated at time $t+1$ using the following map:\begin{equation}
\mathbf{x}(t+1)=f\left(\mathbf{w},\mathbf{x}(t)\right),\end{equation}
 where\begin{equation}
f\left(\mathbf{w},\mathbf{x}(t)\right)=[f_{1}\left(\mathbf{w},\mathbf{x}(t)\right),...,f_{N}\left(\mathbf{w},\mathbf{x}(t)\right)]^{T},\end{equation}
 and

\begin{equation}
f_{n}\left(\mathbf{w},\mathbf{x}(t)\right)=\tanh\left(\sum_{m=1}^{N}w_{nm}x_{m}(t)+x_{0}\right),\quad n=1,...,N.\end{equation}
 Here, $\mathbf{w}$ is an $N\times N$ interaction matrix, with the
following randomly assigned elements: \begin{equation}
w_{nm}=\left\{ \begin{array}{ccc}
-1 & with\: probability & \frac{k}{2N}\\
0 & with\: probability & \frac{N-k}{N}\\
+1 & with\: probability & \frac{k}{2N}\end{array}\right.,\end{equation}
 and $k$ is the average in-degree of the network.

The interaction weights can be interpreted as excitatory, if $w_{nm}=1$,
and respectively inhibitory, if $w_{nm}=-1$. Also, we have $w_{nm}=0$,
if $x_{m}$ is not an input to $x_{n}$. Obviously, the threshold
$x_{0}$ can be considered as a constant input, with a fixed weight
$w_{n0}=1$, to each variable $x_{n}$. Therefore, in the following
discussion we do not lose generality by assuming that the threshold
parameter is always set to $x_{0}=0$.

\section{Phase transition }

In order to illustrate the complex dynamics of the NLRN system, we
consider the results of the simulation of three networks, each containing
$N=128$ variables, and having different average in-degrees: $k=1$,
$k=2$ and respectively $k=4$. Also, the continuous values of the
variables $x_{n}(t)$ are encoded in shades of gray, with black and
white corresponding to the extreme values $\pm1$. In Figure 1, one
can easily see the three qualitatively different types of behavior:
ordered $(k=1)$, critical $(k=2)$, and respectively chaotic $(k=4)$.

A quantitative characterization of the transition from the ordered
phase to the chaotic phase is given by the Lyapunov exponents \cite{key-5},
which measure the rate of separation of infinitesimally close trajectories
of a dynamical system. The linearized dynamics in tangent space is
given by:\begin{equation}
\mathbf{\delta x}(t+1)=\mathbf{J}\left(\mathbf{w},\mathbf{x}(t)\right)\delta\mathbf{x}(t),\end{equation}
 where $\mathbf{J}$ is the Jacobian of the map $f$, with the elements\begin{equation}
J_{nm}=\frac{\partial f_{n}}{\partial x_{m}}=w_{nm}\left[1-\tanh^{2}\left(\sum_{m=1}^{N}w_{nm}x_{m}(t)\right)\right],\end{equation}
 and $\delta\mathbf{x}(t)$ is the separation vector. The dynamics
of $\delta\mathbf{x}(t)$ is typically very complex, involving rotation
and stretching. Therefore, the rate of separation can be different
for different orientations of initial separation vector, such that
one obtains a whole spectrum of Lyapunov exponents. In general, there
are $N$ possible values, which can be ordered: $\lambda_{1}\geq\lambda_{2}\geq...\geq\lambda_{N}$.
These Lyapunov exponents are associated with the Lyapunov vectors,
$v_{1},v_{2},...,v_{N}$, which form a basis in the tangent space.
A perturbation along $v_{n}$ will grow exponentially with a rate
$\lambda_{n}$. Oseledec's theorem \cite{key-6} proves that the following
limit exists:\begin{equation}
\lambda=\lim_{t\rightarrow\infty}\frac{1}{t}\ln\frac{\left\Vert \delta\mathbf{x}(t)\right\Vert }{\left\Vert \delta\mathbf{x}(0)\right\Vert }.\end{equation}
 We should note that, Oseledec's limit will always correspond to $\lambda_{1}$,
because an initial random perturbation will always have a component
along the most unstable direction, $v_{1}$, and because the exponential
growth rate the effect of the other exponents will be obliterated
over time. Thus, in general, it is enough to consider only the maximal
Lyapunov exponent (MLE), which is enough to characterize the behavior
of the dynamical system \cite{key-5}. A negative MLE corresponds to an
ordered system (fixed points and periodic dynamics), while a positive MLE 
is an indication that the system is chaotic. A zero MLE is associated with 
quasiperiodic dynamics and corresponds to the critical transition. 
Figure 2 shows the MLE as a function of the average in-degree,
$\lambda(k)$. One can see that the critical in-degree is $k_{c}=2$,
such that for $k<k_{c}$ the NLRNs are ordered, and for $k>k_{c}$
the NLRs become chaotic. The numerical results were obtained by averaging
over the NLRNs ensemble for each $k$, using $M=256$ NLRNs with $N=256$
elements. Also, for each time series we have discarded the first $1024$
steps, in order to eliminate the transient, and the MLE was calculated
from the next $1024$ steps.

In order to provide a more detailed characterization of the order-chaos
phase transition we introduce the following spectral complexity measure:
\begin{equation}
Q_{\omega}=H_{\omega}D_{\omega},\end{equation}
 where $H_{\omega}$ is the spectral entropy, and $D_{\omega}$ is
the spectral disequilibrium. The complexity is defined by the interplay
of two antagonistic behaviors: the increase of entropy as the system
becomes more and more disordered and the decrease in the disequilibrium
as the system approaches chaos (equiprobability). A similar complexity
measure, evaluated in the direct (time) space, was introduced in \cite{key-7},
for discrete state systems. In contrast, our complexity measure is
defined for continuous state systems, and it is evaluated in the inverse
(frequency) space.

In order to define the spectral entropy \cite{key-8}, we consider the discrete
Fourier transform (DFT): \begin{equation}
\mathbf{X}_{n}(\omega)=F_{\omega}[\mathbf{x}_{n}(t)]=[X_{n}(1),...,X_{n}(\Omega)]^{T},\end{equation}
 \begin{equation}
X_{n}(\omega)=\sum_{t=1}^{T}x_{n}(t)\exp(-2\pi i\omega t/T),\quad\omega=1,...,\Omega,\end{equation}
 and the power spectrum: \begin{equation}
\mathbf{Y}_{n}(\omega)=[Y_{n}(1),...,Y_{n}(\Omega)]^{T},\end{equation}
 \begin{equation}
Y_{n}(\omega)=X_{n}^{*}(\omega)X_{n}(\omega)=\left|X_{n}(\omega)\right|^{2},\quad\omega=1,...,\Omega,\end{equation}
 of the time series: \begin{equation}
\mathbf{x}_{n}(t)=[x_{n}(1),...,x_{n}(T)]^{T},\end{equation}
 corresponding to the attractor of the variable $n$ of a given NLRN.
Here, $X_{n}^{*}$ stands for the complex conjugate value. Since the
variables $x_{n}(t)$ are real, the DFT result has the following symmetry:\begin{equation}
X_{n}(\omega)=X_{n}^{*}(T-\omega),\end{equation}
 and therefore the power spectrum $\mathbf{Y}_{n}(\omega)$ has only
$\Omega=T/2$ positive values:

One can normalize the power spectrum such that:\begin{equation}
p_{n}(\omega)=\frac{Y_{n}(\omega)}{\sum_{\omega=1}^{\Omega}Y_{n}(\omega)},\quad\omega=1,...,\Omega,\end{equation}
 and\begin{equation}
\sum_{\omega=1}^{\Omega}p_{n}(\omega)=1.\end{equation}
 The new variable $p_{n}(\omega)$ can be interpreted as the probability
of having the frequency $\omega$ \textit{embedded} in the time series
$\mathbf{x}_{n}(t)$. Thus, using the spectral probability vector
\begin{equation}
\mathbf{p}_{n}(\omega)=[p_{n}(1),...,p_{n}(\Omega)]^{T},\end{equation}
 one can define the spectral entropy of the time series $\mathbf{x}_{n}(t)$,
as following:\begin{equation}
H_{\omega}[\mathbf{p}_{n}(\omega)]=-\frac{1}{\log_{2}\Omega}\sum_{\omega=1}^{\Omega}p_{n}(\omega)\log_{2}p_{n}(\omega),\end{equation}
 where $\log_{2}\Omega$ is the normalization constant, such that
$0\leq H_{\omega}\leq1$.

Obviously, the spectral entropy of the ordered systems will be low,
$H_{\omega}\sim0$, since only a very small number of frequencies
are present, while the spectral entropy of chaotic systems will be
high, $H_{\omega}\sim1$, since a large number of frequencies are
present. The spectral entropy takes the maximum value, $H_{\omega}=1$,
for the equilibrium state, which is defined deep in the chaotic regime,
where all frequencies become equiprobable: $p(\omega)=\Omega^{-1}$,
$\omega=1,...,\Omega$.

The spectral disequilibrium of the time series $\mathbf{x}_{n}(t)$,
measures the displacement of the corresponding probability vector
$\mathbf{p}_{n}(\omega)$ from the equilibrium state, and it is defined
as following:\begin{equation}
D_{\omega}[\mathbf{p}_{n}(\omega),\Omega^{-1}]=\sum_{\omega=1}^{\Omega}[p_{n}(\omega)-\Omega^{-1}]^{2}.\end{equation}
 A special attention is necessary in the case when the attractor is
zero: $\mathbf{x}_{n}(t)=0$. In this particular case, the power spectrum
is also zero, $\mathbf{Y}_{n}(\omega)=0$, and the probability vector
$\mathbf{p}_{n}(\omega)$ is undetermined. In order to overcome this
difficulty, we define $H_{\omega}=0$ and $D_{\omega}=1$ for this
particular attractor, such that it has the lowest entropy and the
largest displacement from equilibrium.

Since the spectral disequilibrium measures the distance between two
distributions, one may consider also the spectral Kullback-Leibler
divergence \cite{key-9} as an alternative. However, for the considered NLRN model,
the Kullback-Leibler divergence is simply given by: \begin{equation}
D_{\omega}^{KL}[\mathbf{p}_{n}(\omega)||\Omega^{-1}]=\frac{1}{\log_{2}\Omega}\sum_{\omega=1}^{\Omega}p_{n}(\omega)\log_{2}\left(\frac{p_{n}(\omega)}{\Omega^{-1}}\right)=1-H_{\omega}[\mathbf{p}_{n}(\omega)].\end{equation}
Similarly, one can show that the symmetrical Kullback-Leibler divergence
is given by:\begin{equation}
D_{\omega}^{KL}[\mathbf{p}_{n}(\omega)||\Omega^{-1}]+D_{\omega}^{KL}[\Omega^{-1}||\mathbf{p}_{n}(\omega)]=-H_{\omega}[\mathbf{p}_{n}(\omega)]-\frac{1}{\Omega\log_{2}\Omega}\sum_{\omega=1}^{\Omega}\log_{2}p_{n}(\omega).\end{equation}
Therefore, in this case, the Kullback-Leibler divergence
(or its symmetrical version) can be expressed in terms of entropy. Thus, the
spectral disequilibrium seems to be a more appropriate distance measure,
since it cannot be expressed in terms of entropy.

Another quantity of interest is the pairwise spectral correlation
between the power spectrum of two network variables $n$ and $m$,
which is defined as:\begin{equation}
C_{\omega}[\mathbf{Y}_{n},\mathbf{Y}_{m}]=\frac{(\mathbf{Y}_{n}-\mathbf{\overline{Y}}_{n})^{T}\mathbf{(Y}_{m}-\mathbf{\overline{\mathbf{Y}}_{m}})}{\left\Vert \mathbf{Y}_{n}-\mathbf{\overline{Y}}_{n}\right\Vert \left\Vert \mathbf{Y}_{m}-\overline{\mathbf{Y}}_{m}\right\Vert },\end{equation}
 where $\mathbf{\overline{Y}}_{n}$ and $\mathbf{\overline{Y}}_{m}$
represents the mean values.

The average correlation for a given NLRN is: \begin{equation}
C_{\omega}=\frac{1}{N(N-1)}\sum_{n=1}^{N}\sum_{m=1}^{N}[1-\delta(n,m)]C_{\omega}[\mathbf{Y}_{n},\mathbf{Y}_{m}],\end{equation}
 where we have excluded the self-correlation terms ($\delta(n,m)=1$
if $m=n$ and $\delta(n,m)=0$ if $m\neq n$).

In Figure 3 we give the numerical results for the above spectral measures
(entropy, disequilibrium, complexity and correlation), obtained by
averaging over the NLRNs ensemble ($M=256$ networks with $N=256$
elements and $T=1024$). One can see that both the complexity and
the correlation measures are maximized by the critical NLRNs with
$k_{c}=2$.

As mentioned at the beginning of the paper, the continuous NLRN model
is directly related to the binary RTN model, which has been extensively
studied recently \cite{key-4}, \cite{key-10}. Recently, we have investigated
the binary RTN model, using similar quantities, complexity, entropy,
and the mutual information, which are well defined in the time domain.
The obtained results for both NLRN and RTN models are in very good
agreement, showing a phase transition for the same critical connectivity
$k_{c}=2$. Also, for the RTN model, we have shown that the mutual
information, which is the binary counter part of the spectral correlation,
is maximized for $k_{c}=2$. Similar results have also been previously
reported for the RBN model \cite{key-3}.

\section{Conclusion}

We have shown numerically that the NLRN model exhibits an order-chaos
phase transition, for the same critical connectivity value $k_{c}=2$,
as the RBN and RTN models. Also, we have shown that both the pairwise
correlation and the complexity measures are maximized in dynamically
critical networks. These results are in very good agreement with the
previously reported studies on the RBN and RTN models, and show once
again that critical networks provide an optimal coordination of diverse
behavior. We would like also to note that these optimal properties
of critical networks are likely to play a major role in biological
systems, perhaps serving as important selective traits. Given the
potential biological implications, it is of interest that recent data
suggest that genetic regulatory networks in eukaryotic cells are dynamically
critical \cite{key-11}. Also, recent experiments conducted on rat brain
slices show that these neural tissues are critical \cite{key-12}. Thus,
it seems plausible that in cells, neural systems, and other tissues,
natural selection will have acted to maximize both the correlation
across the network, and the diversity of complex behaviors that can
be coordinated within a causal network. Ordered networks have convergent
trajectories, and hence forget their past. Chaotic networks show sensitivity
to initial conditions, and thus they too forget their past, and are
unable to act reliably. On the other hand, critical networks, with
trajectories that on average neither diverge or converge (quasiperiodic dynamics), 
seem best able to bind past to future, and therefore to maximize the correlated
complex behavior.

\pagebreak

\begin{figure}
\centering \includegraphics{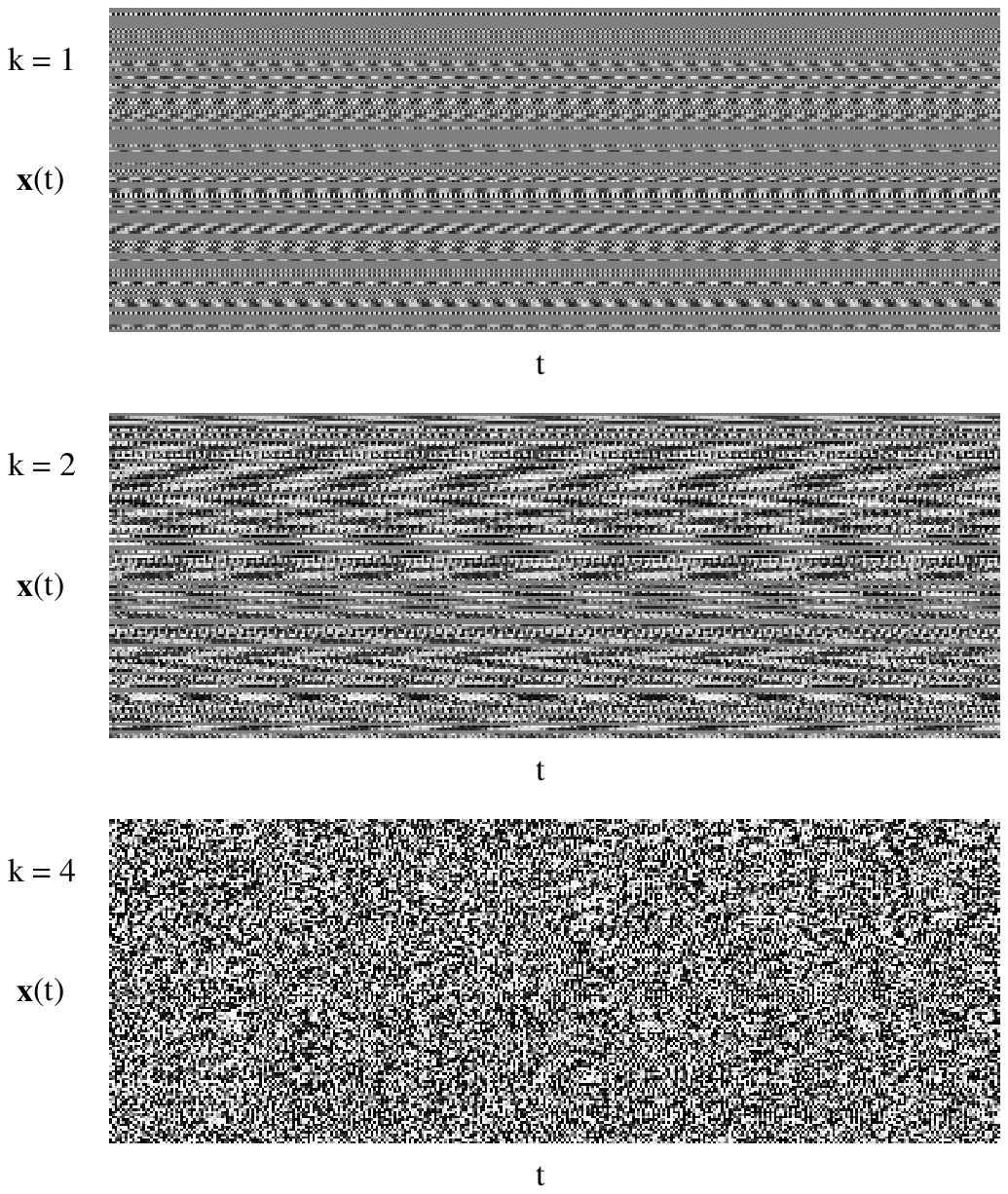} \caption{\label{Fig1} Three qualitatively different types of behavior or the
NLRN model: ordered $(k=1)$, critical $(k=2)$, and respectively
chaotic $(k=4)$.}

\end{figure}

\pagebreak

\begin{figure}
\centering \includegraphics{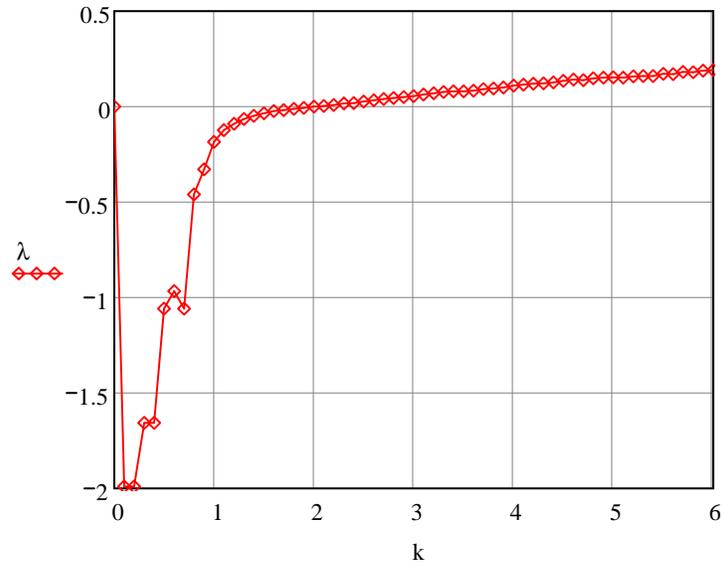} \caption{\label{Fig2} The maximal Lyapunov exponent of the NLRN model, as
a function of the connectivity: $\lambda(k)$.}

\end{figure}

\begin{figure}
\centering \includegraphics{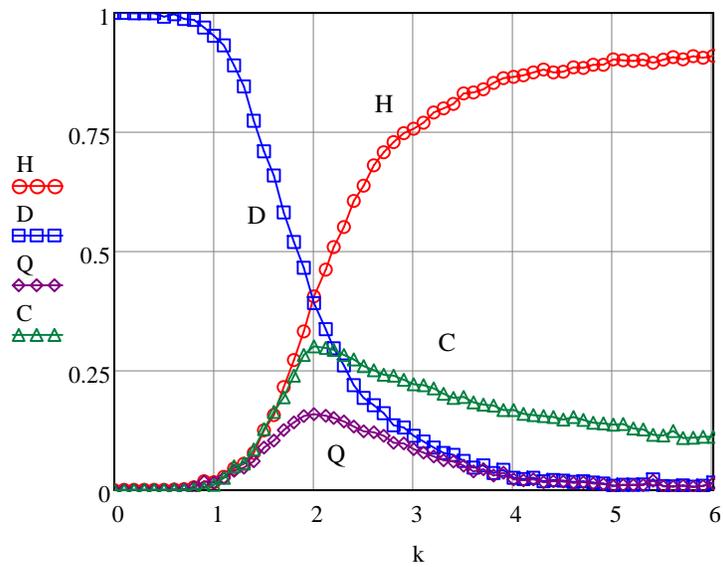} \caption{\label{Fig3} The spectral measures of the NLRN model as function
of of connectivity: entropy $H_{\omega}(k)$, disequilibrium $D_{\omega}(k)$,
complexity $Q_{\omega}(k)$, and correlation $C_{\omega}(k)$.}

\end{figure}

\end{document}